\documentclass[useAMS,usenatbib,a4paper]{mn2e}
 
\usepackage{graphicx, epsfig}
\usepackage{float}
\usepackage{subfig}
\usepackage{natbib}
\usepackage{color}
\usepackage{cancel}
\usepackage{mathrsfs}
\usepackage {amsbsy,amssymb,amsmath}

\newcommand{\be}{\begin{equation}}
\newcommand{\ee}{\end{equation}}
\newcommand{\bea}{\begin{eqnarray}}
\newcommand{\eea}{\end{eqnarray}}

\newcommand{\hatchi}{\hat{\chi}}

\begin{document}

\title[Prospects for clustering and lensing measurements using IM]
{Prospects for clustering and lensing measurements with forthcoming intensity mapping and optical surveys}
\author[ Pourtsidou, Bacon, Crittenden \& Metcalf ]{ A. Pourtsidou$^{1}$, D. Bacon$^{1}$, R. Crittenden$^{1}$, R. B. Metcalf$^{\, 2}$  \\ 
$^{1}$ Institute of Cosmology \& Gravitation, University of Portsmouth, Burnaby Road, Portsmouth, PO1 3FX, United Kingdom \\
$^2$ Dipartimento di Fisica e Astronomia, Alma Mater Studiorum Universit\'{a} di Bologna, viale Berti Pichat, 6/2, I-40127, Bologna, Italy
}
%
%\date{Accepted ???, Received ???; in original form \today}
%\pagerange{\pageref{firstpage}--\pageref{lastpage}}
%\pubyear{2004}

\maketitle

\begin{abstract}
We explore the potential of using intensity mapping surveys (MeerKAT, SKA) and optical galaxy surveys (DES, LSST) to detect HI clustering and weak gravitational lensing of 21 cm emission in auto- and cross-correlation. Our forecasts show that high precision measurements of the clustering and lensing signals can be made in the near future using the intensity mapping technique. Such studies can be used to test the intensity mapping method, and constrain parameters such as the HI density $\Omega_{\rm HI}$, the HI bias $b_{\rm HI}$ and the galaxy-HI correlation coefficient $r_{\rm HI-g}$.
\end{abstract}


\maketitle

\section{Introduction}

Intensity mapping \citep{Battye:2004re,Chang:2007xk,Loeb:2008hg,Mao:2008ug,Peterson:2009ka,Seo:2009fq,Ansari:2011bv,Battye:2012tg,Switzer:2013ewa,Bull:2014rha} is an innovative technique that uses neutral hydrogen (HI) to map the large-scale structure of the Universe in three dimensions. Instead of detecting individual galaxies like the conventional galaxy surveys, intensity mapping surveys use HI as a dark matter tracer by measuring the intensity of the redshifted 21 cm line across the sky and along redshift, treating the 21 cm sky as a diffuse background, similar to the Cosmic Microwave Background (CMB).  

\citet{Santos:2015bsa} investigated the potential of the planned Square Kilometre Array\footnote{www.skatelescope.org} (SKA) to deliver HI intensity mapping maps over a broad range of frequencies and a substantial fraction of the sky. Detecting the 21 cm signal in auto- and cross-correlation using intensity mapping and optical galaxy surveys is essential in order to exploit the intensity mapping technique, test foreground removal methods, and identify and control systematic effects. This is possible using SKA pathfinders like MeerKAT\footnote{http://www.ska.ac.za/meerkat/} and, as we will show, in many cases high signal-to-noise ratio measurements can be achieved.

Cross-correlation between large scale structure (LSS) traced by galaxies and 21 cm intensity maps at $z \sim 1$ was first detected using the Green Bank Telescope (GBT) and the DEEP2 optical galaxy redshift survey \citep{Chang:2010jp}; this measurement was improved using the GBT and the WiggleZ Dark Energy Survey \citep{Masui:2012zc}.  The auto-power spectrum of 21 cm intensity fluctuations using data acquired with the GBT was used in \citet{Switzer:2013ewa} to constrain HI fluctuations at $z\sim 0.8$ and was interpreted as an upper bound on the 21 cm signal because of residual foreground contamination bias.

In this work we present HI detection forecasts for auto- and cross-correlation measurements using intensity mapping surveys with MeerKAT and SKA, and optical galaxy surveys with the Dark Energy Survey (DES)\footnote{http://www.darkenergysurvey.org/} and the Large Synoptic Survey Telescope (LSST)\footnote{http://www.lsst.org/}. Our forecasts concern both the HI intensity fluctuations as well as the weak gravitational lensing of 21 cm emission, using the weak lensing intensity mapping method developed in \citet{Pourtsidou:2013hea,Pourtsidou:2014pra}. 
In the following we denote the density fluctuations $\delta$ using the subscript HI for 21 cm and $g$ for galaxies. We also denote the lensing convergence $\kappa$ using the subscript $g$ when it is detected using galaxies and IM when using the intensity mapping method.

In Section~\ref{sec:surveys} we introduce the HI intensity mapping and optical galaxy surveys we are going to use for our clustering and lensing measurements forecasts and analyse their noise properties.
In Section~\ref{sec:auto} we study correlations of the HI observables. We investigate the possibility of measuring the HI-HI power spectrum ($\delta_{\rm HI} \times \delta_{\rm HI}$) with MeerKAT and show forecasts for the lensing convergence power spectrum measurements ($\kappa_{\rm IM} \times \kappa_{\rm IM}$) and for $\delta_{\rm HI} \times \kappa_{\rm IM}$ using MeerKAT/SKA Phase 1 (SKA1) and the intensity mapping method.  
Cross-correlation studies are less susceptible to systematic contamination than auto-correlations, and can be observed when the noise levels in the HI observations are relatively high.  
We study these in Section~\ref{sec:cross}. First we examine the possibility of measuring the $\delta_{\rm HI} \times \delta_g$ and $\delta_{\rm HI} \times \kappa_g$ correlations using MeerKAT and DES.    
We then study the $\delta_g \times \kappa_{\rm IM}$ correlation with LSST and MeerKAT/SKA1. Finally, we investigate $\kappa_g \times \kappa_{\rm IM}$ with LSST and SKA1.

There are exciting prospects for performing clustering and lensing measurements with the forthcoming intensity mapping and optical surveys. The signal-to-noise ratio for many of the cross and auto-spectra we consider is high, so significant progress will occur in the near future, exploiting SKA pathfinders and near-term optical galaxy surveys. The primary goal of our work is to show that it is possible to perform high precision clustering and lensing studies in three dimensions using the intensity mapping technique. We can use these measurements to calibrate the neutral gas density $\Omega_{\rm HI}$, the HI bias parameter $b_{\rm HI}$ and the galaxy-HI correlation coefficient $r_{\rm HI-g}$. 
{\color{black}  The current uncertainties in the HI density fraction $\Omega_{\rm HI}$ and the bias $b_{\rm HI}$ are large; for example, the best constraint obtained so far for the HI density - HI bias combination is $\Omega_{\rm HI}b_{\rm HI}=4.3 \pm 1.1 \, \times 10^{-4}$ at $z\sim 0.8$ \citep{Switzer:2013ewa}. Their precise values and evolution across redshift are very important for the signal-to-noise ratio of the clustering and lensing measurements, as they determine the amplitude of the HI signal. As shown in \citet{Bull:2014rha}, $\Omega_{\rm HI}(z)$ substantially affects the forecasted cosmological constraints using late-time intensity mapping clustering surveys, and it is also very important for the 21cm lensing signal-to-noise ratio from post-reionization source redshifts \citep{Pourtsidou:2014pra}. Therefore, it is crucial to utilise near term intensity mapping surveys in order to tightly constrain them. Forecasted constraints on the HI parameters and other cosmological parameters using clustering and lensing measurements will be the subject of future work. }

\section{The Surveys}
\label{sec:surveys}

\subsection{HI intensity mapping}

We consider a range of HI surveys, focussing on the SKA and its pathfinder MeerKAT. There are two different observing modes we can consider, namely the single-dish mode and the interferometer mode (see \citet{Bull:2014rha} for details). Below we describe the noise properties for both modes. 

\subsubsection{Single-dish mode}

{\color{black} The SKA-MID instrument is primarily an interferometer, but there are also discussions and plans to operate it in single-dish mode as well, in order to collect total power (auto-correlation) data \citep{ECPauto, Bull:2014rha, Santos:2015bsa}. This is crucial for cosmological measurements with the SKA an{blackd the HI intensity mapping technique. For example, arrays with large dishes do not adequately sample BAO scales at low redshifts in interferometer mode, as the largest scale probed is limited by the dish size. Using the single-dish mode and covering a large fraction of the sky ultra large scales can be measured and the constraints obtained are competitive with state-of-the-art optical galaxy surveys like Euclid \citep{Amendola:2012ys, Santos:2015bsa}.}

MeerKAT is a 64-dish SKA pathfinder on the planned site of SKA1-MID and it will start observing in 2016 with at least 16 dishes. From here onwards, we will refer to its first phase as MeerKAT-16, and its full phase as MeerKAT. The dishes have $13.5 \, {\rm m}$ diameter with number of beams $N_{\rm beams}=1$; the redshift {\color{black} (frequency)} range is $0<z<1.45$ {\color{black}($580<f<1420$ MHz) }for the 21 cm line and the frequency resolution $\Delta f = 50  \, {\rm kHz}$. The system temperature is taken to be $T_{\rm sys}=25 \, {\rm K}$. The sky area and total observing time are determined by the survey strategy. We will consider two strategies: First, we assume a sky area $A_{\rm sky}=1000 \, {\rm deg}^2$ and a total observation time of $3$ weeks, and then we repeat the calculation with $A_{\rm sky}=5000 \, {\rm deg}^2$ and a total observation time of $15$ weeks.  

The noise properties of such measurements have been described in various works (see, for example, \citet{Battye:2012tg}) and depend on the instrumental noise in a given pixel (beam), its volume, and the instrumental response, modelled by the window function $W(k)$. 
Because the frequency resolution in such surveys is very good (of the order of tens of kHz) we can ignore the instrument response function in the radial direction. However, there is a window function related to the  finite angular resolution:
\be
W^2(k)={\rm exp}\left[-k^2\chi(z)^2\left(\frac{\theta_{\rm B}}{\sqrt{8{\rm ln}2}}\right)^2\right],
\ee where $\chi(z)$ is the comoving radial distance at redshift $z$ and $\theta_{\rm B} \sim \lambda/D_{\rm dish}$ the beam FWHM of a single dish with diameter $D_{\rm dish}$ at wavelength $\lambda$. 
Considering a redshift bin with limits $z_{\rm min}$ and $z_{\rm max}$, the survey volume will be given by
\be
V_{\rm sur}=\Omega_{\rm tot} \int_{z_{\rm min}}^{z_{\rm max}} dz \frac{dV}{dz d\Omega} =  \Omega_{\rm tot} \int_{z_{\rm min}}^{z_{\rm max}} dz \frac{c \chi(z)^2}{H(z)},
\label{eq:volume}
\ee and $ \Omega_{\rm tot} = A_{\rm sky}$, the sky area the survey scans. The pixel's volume $V_{\rm pix}$ is also calculated from Eq.~(\ref{eq:volume}), but with 
\be
 \Omega_{\rm pix} \simeq 1.13\theta^2_B
\ee assuming a Gaussian beam, and the corresponding pixel $z$-limits corresponding to the channel width $\Delta f$. Finally, the pixel noise $\sigma_{\rm pix}$ is given by
\be
\sigma_{\rm pix}=\frac{T_{\rm sys}}{\sqrt{\Delta f \, t_{\rm total}(\Omega_{\rm pix}/\Omega_{\rm tot})N_{\rm dishes} N_{\rm beams}}},
\ee with $N_{\rm dishes}$ the number of dishes.

{\color{black} Here we should note that various systematic effects might lead to an increase of the actual noise in the autocorrelation measurements. For example, in addition to the thermal noise quantified above these observations suffer from correlated $(1/f)$ noise and ground pickup. However, recent work on the subject suggests that these effects can be removed to a large extent \citep{Bigot-Sazy:2015jaa}. Other systematics include real beams with sidelobes and mis-calibration which will lead to mode-mixing and thus affect the foreground subtraction.}

The MeerKAT radio telescope is a precursor to the SKA telescope and will be integrated into the mid-frequency component of SKA1 (SKA1-MID).  As we will see below, MeerKAT can also be used as an interferometer in its own right.

\subsubsection{Interferometer mode}

{\color{black} 
The thermal noise power spectrum for an interferometer array is given by \citep{White:1999uk, Zaldarriaga:2003du}
\be
C^{\rm N}_\ell = \frac{(2\pi)^2T^2_{\rm sys}}{Bt_u d^2\ell},
\ee where $B$ is the total bandwidth of the observation, $t_u$ is the time each visibility is observed, and $\ell$ is related to the Fourier wavenumber $u$ by $\ell=2\pi u$ -- {\color{black} consequently, the Fourier space pixel $d^2\ell$ is related to the square resolution element $d^2u$ by $d^2\ell=(2\pi)^2d^2u$}. 
The observation time per visibility $t_u$ is given by \citep{Zaldarriaga:2003du, Mao:2008ug}
\be
t_u =  \frac{A_{\rm dish}}{\lambda^2} t_0 n(u),
\ee where $A_{\rm dish}$ is the area of an individual dish, $t_0$ is the total observation time and $n(u)$ -or, equivalently, $n(\ell)$- is the number density of baselines. 

Using the above we finally get 
\be
C^{\rm N}_\ell = \frac{T^2_{\rm sys}[{\rm FOV}]^2}{Bt_0 n(\ell)}.
\label{eq:CellN}
\ee
{\color{black} Here we have used the fact that the primary beam size (and hence $d^2u$) is related to the area of the dishes, so we can use the approximation $A_{\rm dish}=\lambda^2 d^2u$ \citep{Zaldarriaga:2003du}, and $1/{\rm FOV} \equiv A_{\rm dish}/\lambda^2$ (where FOV is the field of view, and $\lambda$ is the observing wavelength).}
The required $n(\ell)$ distributions to calculate the noise of SKA1-MID and MeerKAT in interferometer mode are taken from \citet{Bull:2014rha}. The system temperature $T_{\rm sys}$ is the sum of the sky and receiver noise and is approximately given by (the $T_{\rm sys}$ values are nominal and depend on the sky and the receivers) \citep{Dewdney13}
\be
T_{\rm sys} = 28 + 66\left(\frac{\nu}{300 \, {\rm MHz}}\right)^{-2.55} \, {\rm K},
\ee with $\nu$ the observing frequency. 
} {\color{black} We also note that using the uniform approximation formula for the number density of baselines in Equation~(\ref{eq:CellN}), $n(\ell) \simeq (2\pi)N^2_{\rm dishes}/\ell^2_{\rm max}$, we recover the widely known uniform $C^{\rm N}_\ell$ formula (see, for example, \citep{Zaldarriaga:2003du}).}

The thermal noise of the interferometer is part of the lensing reconstruction noise using the lensing estimator developed in \citet{Pourtsidou:2014pra}.
In that work the method of 21 cm intensity mapping was used to study gravitational lensing over a wide range of post-reionization redshifts --- this extends weak lensing measurements to higher redshifts than are accessible with conventional galaxy surveys. Detecting $\kappa$ with this method would be an important science achievement of the intensity mapping technique. 

Central to this detection is understanding $N_\kappa(\ell)$, the lensing reconstruction noise using the aforementioned method.
The expression for $N_\kappa(\ell)$ is rather lengthy, so we will not include it here, but the interested reader is referred to  \citet{Pourtsidou:2014pra}, Appendix C. To summarise, the lensing reconstruction noise involves the underlying dark matter power spectrum $P_{\delta \delta}$, the HI density $\Omega_{\rm HI}(z)$ as well as the HI mass (or luminosity) moments up to 4th order and, as already stated, the thermal noise of the instrument $C^{\rm N}_\ell$.  
{\color{black} Note that in the following we will assume an observation (HI source) redshift $z_s=1.4$ corresponding to a frequency of $592 \, {\rm MHz}$, bandwidth $B=40 \, {\rm MHz}$ corresponding to $\Delta z \sim 0.15$, total observation time $t_0=4,000 \, {\rm hrs}$ and sky area $A_{\rm sky}=25,000 \, {\rm deg}^2$ when we consider MeerKAT and SKA1 in interferometer mode. We remind the reader that the frequency (redshift) range for MeerKAT is $580<z<1420$ MHz ($0<z<1.45$), while for SKA1-MID $350<f<1050$ MHz ($0.35<z<3.06$) (Band $1$) \citep{Bull:2014rha}.}

\subsection{Optical galaxy surveys}

We consider two photometric surveys: the ongoing Dark Energy Survey (DES) and the planned Large Synoptic Survey Telescope (LSST).  DES aims to investigate the nature of the cosmic acceleration and combines four probes of Dark Energy, namely Type Ia Supernovae, Baryonic Acoustic Oscillations (BAOs), galaxy clusters and weak gravitational lensing. LSST is a ground based, wide field survey telescope. One of its main goals is to provide multiple probes of dark energy, with the two most powerful being weak gravitational lens tomography and BAOs.

The DES survey parameters are \citep{Becker:2015ilr} $A_{\rm sky}=5000 \, {\rm deg}^2$, number density of galaxies $n_g = 10 \, {\rm arcmin}^{-2}$, redshift range $0<z<2$ with median redshift $z_0=0.7$. 
The LSST survey parameters are assumed to be $f_{\rm sky}=0.5$, number density of galaxies $n_g = 40 \, {\rm arcmin}^{-2}$, redshift range $0<z<2$ with median redshift $z_0=1$ \citep{Abell:2009aa}. 
The redshift distribution for galaxy surveys like DES and LSST (and Euclid) has the form \citep{Becker:2015ilr, Abell:2009aa, Amendola:2012ys}
\be
\frac{dn}{dz} \propto z^\alpha \, {\rm exp}[-(z/z_0)^\eta].
\ee
For our forecasts we will use the common parametrisation $\alpha=2, \eta=3/2$.

For these surveys, the primary noise for density measurements arises from shot noise, with the shot noise contribution given by
\be
P^{\rm shot} = \frac{1}{(N_g/V_{\rm sur})},
\ee with $N_g$ the number of galaxies within the redshift bin under consideration.

These optical surveys can constrain weak lensing via shear measurements.  The noise associated with the estimated weak lensing convergence is given by  
 $\sigma^2_\kappa/\bar{n}_b$, where $\sigma_\kappa$ is the shape noise of each background galaxy and $\bar{n}_b$ is the number density of background galaxies in the chosen source bin. In the following we assume $\sigma_\kappa = 0.3$ \citep{Schmidt:2011qj}.

\section{HI alone}
\label{sec:auto}

In this Section we investigate auto-correlations of the HI observables, and we show that high signal-to-noise HI detection can be achieved with near-future facilities like MeerKAT-16, hence there are very good prospects for testing and using the intensity mapping method very soon. Lensing of 21 cm sources using the intensity mapping method requires more powerful instruments like the SKA, and heavily depends on the HI density evolution with cosmic time.

\subsection{$\delta_{\rm HI} \times \delta_{\rm HI}$ with MeerKAT-16}

The detection of HI in autocorrelation using the intensity mapping method is the primary science goal of an intensity mapping instrument.
The power spectrum of the HI fluctuations, $\delta_{\rm HI}$, is assumed to take the form
\be
P_{\rm HI}(k,z)=\bar{T}(z)^2 b_{\rm HI}(z)^2 P_{\rm \delta \delta}(k,z),
\ee where $P_{\rm \delta \delta}$ is the underlying dark matter density power spectrum and $b_{\rm HI}$ the HI bias.
The mean HI brightness temperature at redshift $z$ is given by \citep{Battye:2012tg}
\be
\bar{T}(z)=180 \, \Omega_{\rm HI}(z)h  \frac{(1+z)^2}{H(z)/H_0} \, {\rm mK}.
\ee  
 For our forecasts here and in the next Sections we will use $b_{\rm HI}(z)$ from \citet{Camera:2013kpa} 
and assume
\be
\Omega_{\rm HI}(z) = 4 \times 10^{-4} (1+z)^{0.6}
\label{eq:omHImodel}
\ee which has been suggested in \citet{Crighton:2015pza}. We also use the fitting formula by \citet{Smith:2002dz} for the nonlinear power spectrum.

The uncertainty on a power spectrum measurement averaged over a radial bin in $k$-space of width $\Delta k$ is \citep{Battye:2012tg}
\be
\delta P_{\rm HI} = \sqrt{2\frac{(2\pi)^3}{V_{\rm sur}}\frac{1}{4\pi k^2 \Delta k}}[P_{\rm HI}+\sigma^2_{\rm pix}V_{\rm pix}W^{-2}], 
\ee  
where the pixel noise, pixel volume and response window function were described in the previous Section. 

The results for HI detection in autocorrelation at a central redshift $z_c=0.1$ with a redshift bin width $\Delta z = 0.2$ using MeerKAT-16 and the two aforementioned survey strategies (three weeks and $A_{\rm sky}=1000 \, {\rm deg}^2$, fifteen weeks and $A_{\rm sky}=5000 \, {\rm deg}^2$ ) are shown in Fig.~\ref{fig:PHIKAT7}, using $\Delta k=0.01 \, {\rm Mpc}^{-1}$. We plot the cumulative signal-to-noise ratio $(S/N)$, defined as
\be
\frac{\rm S}{\rm N} = \sqrt{\sum_k \left(\frac{P_{\rm HI}}{ \delta P_{\rm HI}}\right)^2}.
\label{eq:SovN}
\ee
As can be seen, these measurements are very precise across a wide range of scales and we can use them to calibrate the combination $\Omega_{\rm HI}b_{\rm HI}$. Note that since MeerKAT will cover a wide redshift range $0<z<1.45$, we can use tomography to probe the combination $\Omega_{\rm HI}b_{\rm HI}$ at different redshifts. 

\begin{figure}
\centerline{
\includegraphics[scale=0.6]{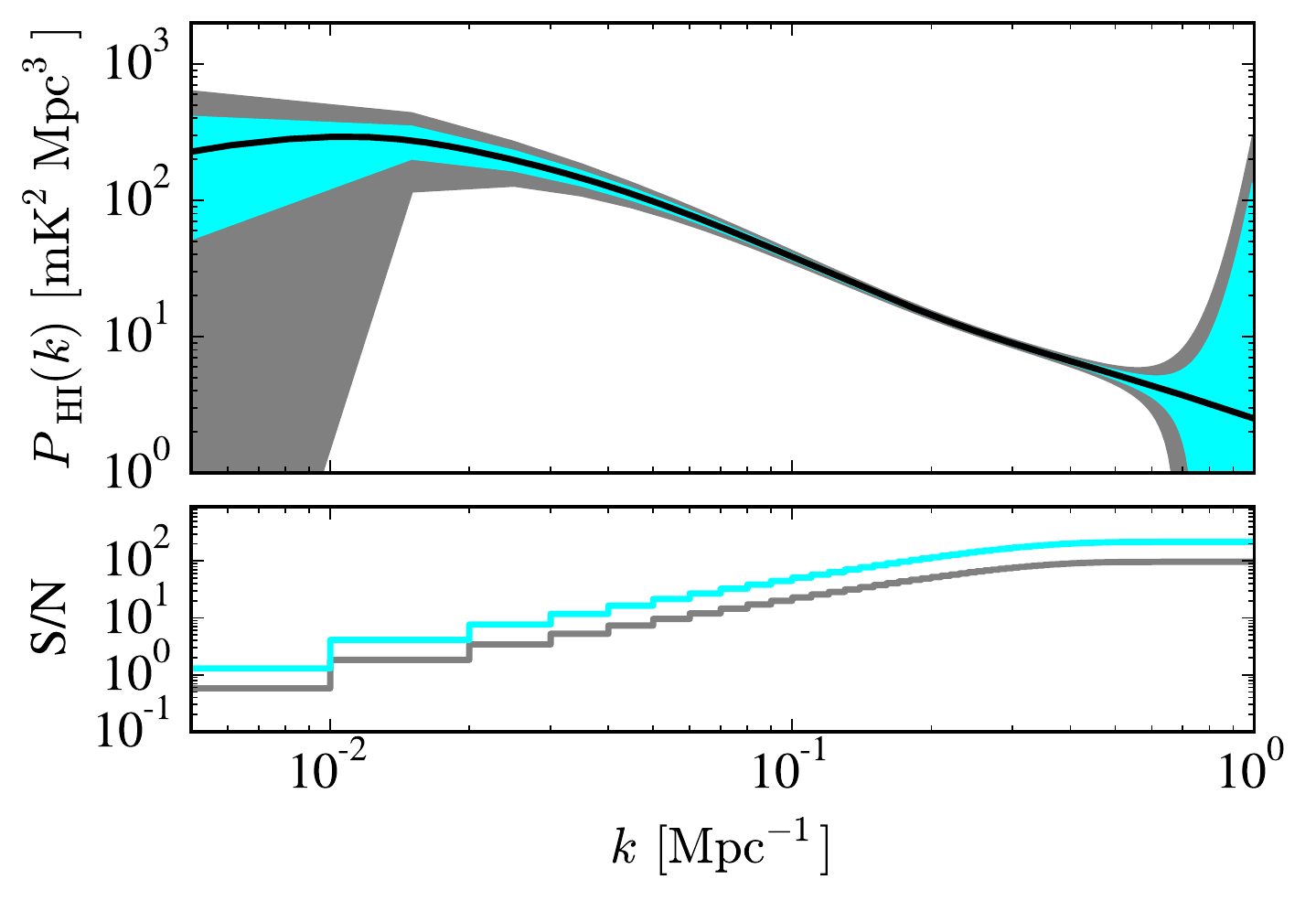}
}
\caption{HI detection in autocorrelation with MeerKAT-16. The upper panel shows the predicted power spectrum $P_{\rm HI}(k,z_c)$ at $z_c=0.1$ (black solid line). The grey area represents the measurement errors $\delta P_{\rm HI}$ taking $A_{\rm sky}=1000 \, {\rm deg}^2$ and a total observation time of $3$ weeks, while the cyan area corresponds to $A_{\rm sky}=5000 \, {\rm deg}^2$ and a total observation time of $15$ weeks. The lower panel shows the cumulative signal-to-noise ratio (S/N) defined in Eq.~(\ref{eq:SovN}).}
\label{fig:PHIKAT7}
\end{figure}

\subsection{$\kappa_{\rm IM} \times \kappa_{\rm IM}$ with MeerKAT/SKA1}

The lensing convergence power spectrum from sources at redshift $z_s$ is given by the expression \citep{Kaiser:1991qi, Kaiser:1996tp} 
\be
C_{\kappa \kappa}(\ell) = \frac{9\Omega^2_m H^3_0}{4c^3} \int^{z_s}_0 dz \frac{P_{\delta \delta}(k=\ell/\chi,z)}{a^2 H(z)/H_0}
\left[\frac{\hatchi_s-\chi}{\hatchi_s}\right]^2,
\ee with $\hatchi_s \equiv \chi(z_s)$.
The uncertainty in the measurement of the power spectrum is
\be
\delta  C_{\kappa \kappa}(\ell) = \sqrt{\frac{2}{(2\ell+1)\Delta \ell f_{\rm sky}}}
\left(C_{\kappa \kappa}(\ell)+N_\kappa(\ell)\right),
\ee where $N_\kappa(\ell)$ is the lensing reconstruction noise using the intensity mapping method described in the previous Section. 

In \citet{Pourtsidou:2014pra} it was found that the signal-to-noise ratio is strongly dependent on the possible evolution of the HI mass function.  
More specifically, it was shown that assuming the no-evolution scenario (which is the most conservative, but also less realistic approach), precise measurements can only be made with an SKA2-like instrument; however assuming instead a model where the HI density $\Omega_{\rm HI}(z)$ increases by a factor of $5$ by redshift $z=3$ and then slowly decreases towards redshift $z=5$, as suggested by the DLA observations from \citet{Peroux:2001ca} {\color{black}(for more recent results in the redshift range $2<z<5$ see \citet{Sanchez-Ramirez:2015adp})}, high signal-to-noise ratio can be achieved even with SKA1. 

In this work we instead use the HI evolution model given by Eq.~(\ref{eq:omHImodel}), which fits observations in a wide redshift range, and we implement this evolution in the $\phi^{\star}$ parameter of the HI mass function which is locally measured by the HIPASS survey \citep{Zwaan:2003hp}. {\color{black} We also note that in \citet{Pourtsidou:2014pra} 
the telescope distribution within the array was approximated as uniform for the calculation of the thermal noise component, while here we use the baseline designs from \citet{Bull:2014rha}.
In Fig.~\ref{fig:CkkSKA} we show results for MeerKAT and SKA1 assuming HI sources are at $z_s=1.4$ and using $\Delta \ell = 50$.}

{\color{black} As we can see, we can detect lensing using the intensity mapping method and SKA1, but using MeerKAT detection in autocorrelation is not possible. However, below we will demonstrate that cross-correlations can enhance the signal-to-noise ratio of the lensing measurements.}
\begin{figure}
\centering
\includegraphics[scale=0.6]{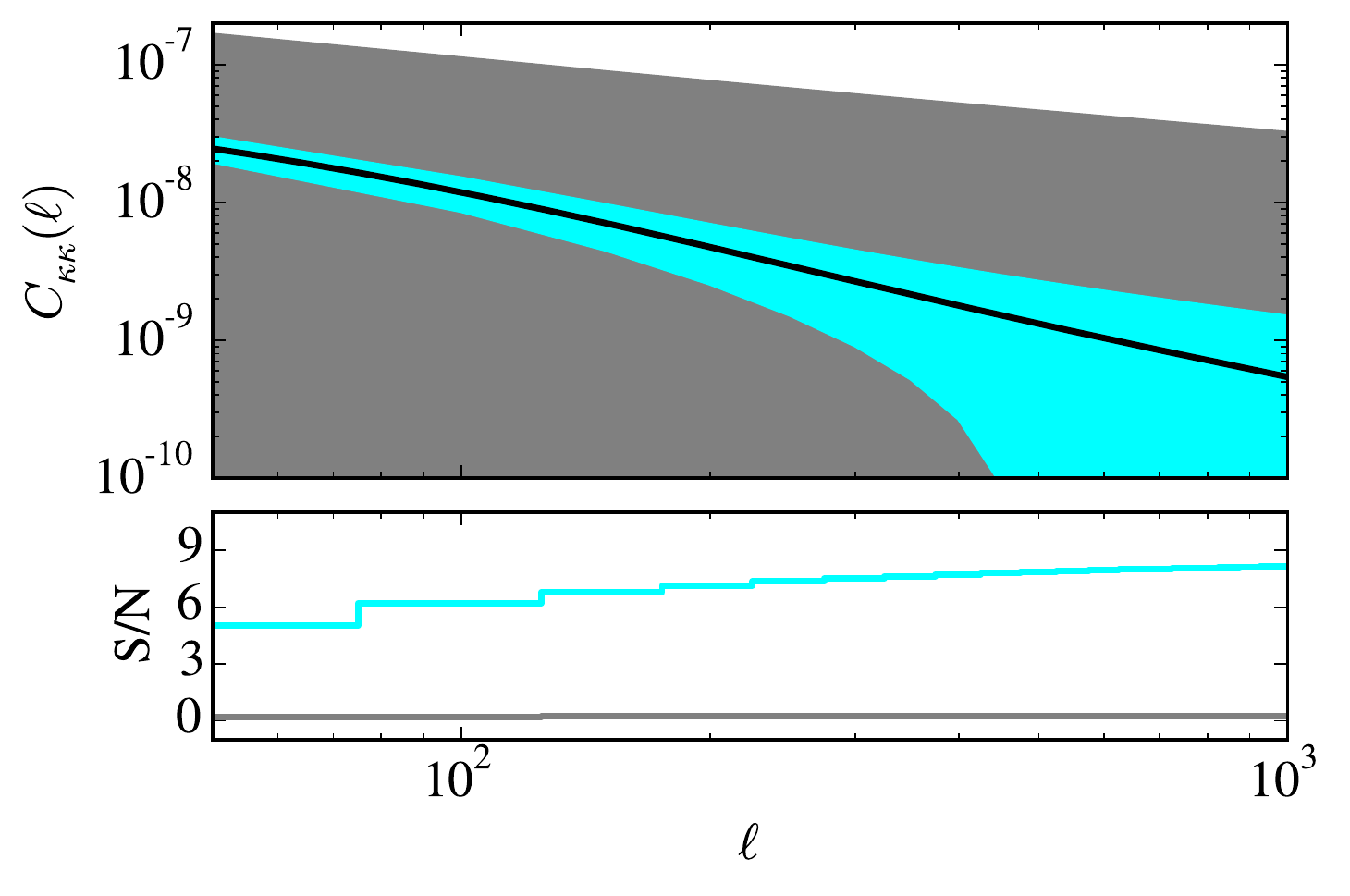}
\caption{The upper panel shows the convergence power spectrum and measurement errors with MeerKAT (grey) and SKA1 (cyan), using the intensity mapping method. The lower panel shows the cumulative signal-to-noise (S/N) ratio. Sources are at $z_s=1.4$.}
\label{fig:CkkSKA}
\end{figure}

\subsection{$\delta_{\rm HI} \times \kappa_{\rm IM}$ with MeerKAT/SKA1}

We are going to examine the correlation of a foreground ($f$) density tracer field with the background ($b$) convergence $\kappa$ field, where both are probed by the IM survey. 

Using the Limber approximation \citep{Limber:1954zz} the angular cross-power spectrum $C_{\rm HI\kappa}$ is given by
\begin{align} \nonumber
C_{\rm HI \kappa}(\ell)&=\frac{3\Omega_{\rm m}H^2_0}{2c^2}\int \frac{d\chi_f}{a(\chi_f)}W_f(\chi_f)
\int d\chi_b W_b(\chi_b) \times \frac{\chi_b-\chi_f}{\chi_b \chi_f} \\ 
&\bar{T}(
\chi_f) r_{\rm HI}b_{\rm HI}(\chi_f) P_{\rm \delta \delta}\left(\frac{\ell}{\chi_f},\chi_f\right),
\end{align} 
where $\chi$ is the comoving distance, $W_f$ ($W_b$) the foreground (background) redshift distribution and $r_{\rm HI}$ is a correlation coefficient quantifying the potential stochasticity between the dark matter density and the HI density fields. If the foreground lens slice is narrow enough in redshift 
($\Delta z \sim 0.1$ is sufficient), 
we can approximate the foreground redshift distribution as a delta function at a distance $\hat{\chi}_f$, $W_f(\chi_f)=\delta^{\rm D}(\chi_f-\hatchi_f)$. We also use the delta function approximation at a distance  $\hat{\chi}_b$ for the distribution of the 21 cm sources.
We then find
\begin{align} \nonumber
C_{\rm HI \kappa}(\ell)=\frac{3\Omega_{\rm m}H^2_0}{2c^2}\left( \frac{\bar{T}(
\hatchi_f) r_{\rm HI}b_{\rm HI}(\hatchi_f) P_{\rm \delta \delta}\left(\frac{\ell}{\hatchi_f},\hatchi_f\right)}{a(\hatchi_f)\hatchi_f}\right)  \frac{\chi_b-\hatchi_f}{\chi_b } .
\end{align} 

It is useful to translate the HI power into multipole space, 
\be
C_{\rm HI-HI}(\ell) = \int dz E(z) W^2(z) [\bar{T}(z)]^2P_{\delta \delta}(\ell/\chi(z),z)/\chi^2(z),
\ee with $W(z)$ a projection kernel which we take to be a top-hat function equal to $1/\Delta z$ within the redshift bin and $0$ otherwise. 

The uncertainty in the cross correlation, for a bin of width $\Delta \ell$ and for a survey scanning a fraction of the sky $f_{\rm sky}$, is
\begin{align} \nonumber
&\delta C_{\rm HI \kappa}(\ell) = \sqrt{\frac{2}{(2\ell+1)\Delta \ell f_{\rm sky}}} \times \\
&\sqrt{C^2_{\rm HI\kappa}(\ell)+\left(C_{\rm HI-HI}(\ell)+N(\ell)\right)\left(C_{\kappa \kappa}(\ell)+N_\kappa(\ell)\right)},
\end{align} with $N_\kappa(\ell)$ from \citet{Pourtsidou:2013hea,Pourtsidou:2014pra}. For the single-dish mode, the noise term $N(\ell)$ is given by \citep{Battye:2012tg}
\be
N(\ell) = \Omega_{\rm pix}(\sigma_{\rm pix})^2 {\rm exp}[\ell(\ell+1)(\theta_B/\sqrt{8{\rm ln}2})^2],
\label{eq:Nelldish}
\ee with $\sigma_{\rm pix}=T_{\rm sys}/\sqrt{2\Delta f t_{\rm obs}}$ (the $1/\sqrt{2}$ factor comes from assuming dual polarisation). For the interferometer mode,
$N(\ell) = C^{\rm N}_\ell$, defined in Eq.~(\ref{eq:CellN}).

The results are shown in Fig.~\ref{fig:CHIkIM} assuming the MeerKAT and SKA1 parameters in interferometer mode. The foreground central redshift is $z_c = 0.5$ with $\Delta z = 0.1$. We see that using MeerKAT in interferometer mode we have the possibility of detecting the lensing convergence in cross correlation with the HI density using the intensity mapping method (with a cumulative ${\rm S/N} \sim 5$). With SKA1 we can achieve a high signal-to-noise ratio detection. Using tomography (for example, taking different foreground bins $z_c$) we can perform measurements with a similar signal-to-noise ratio level along the redshift (frequency) direction.
 
\begin{figure}
\centerline{
\includegraphics[scale=0.6]{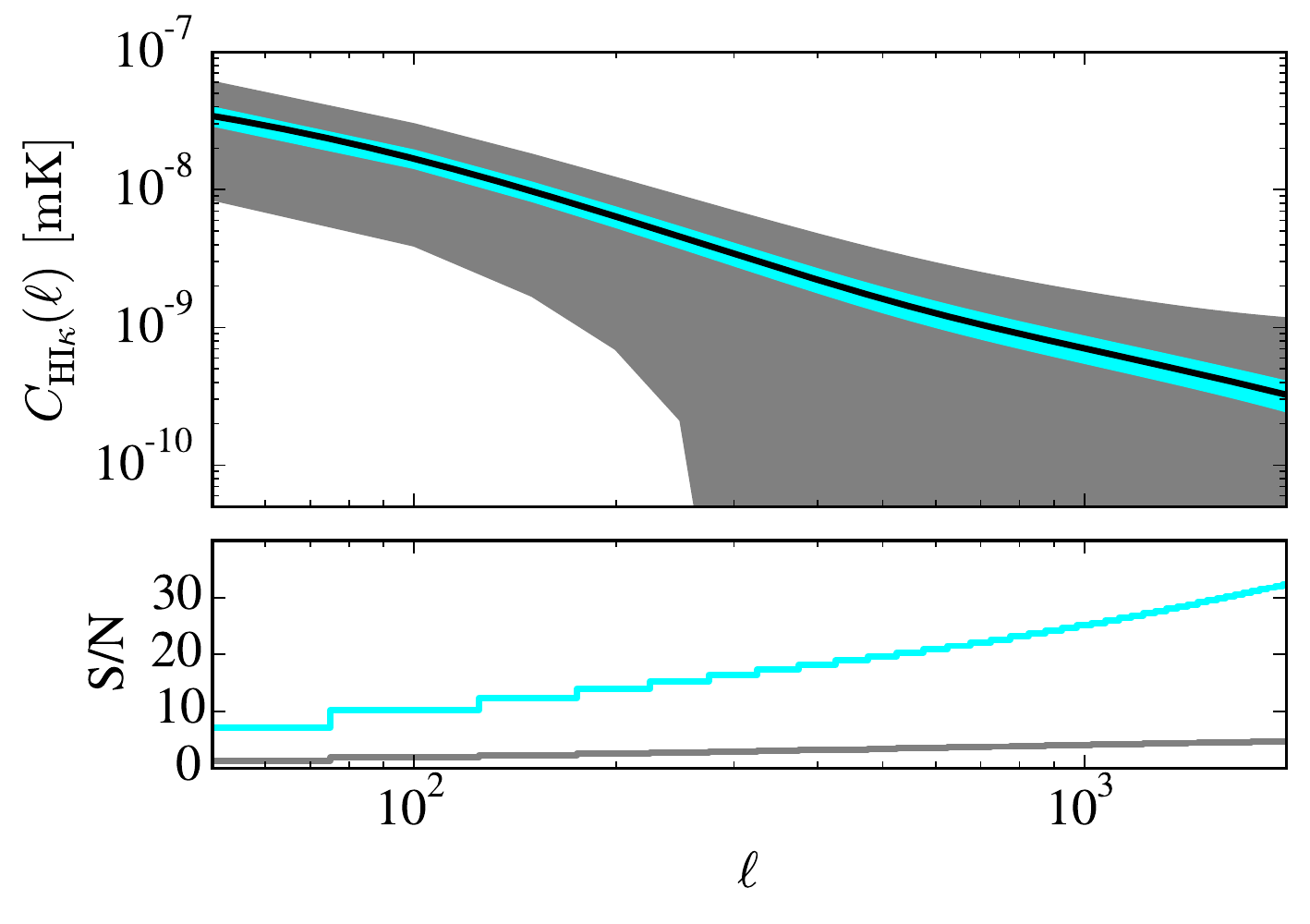}
}
\caption{The upper panel shows the $C_{\rm{HI}\kappa}$ cross correlation power spectrum and measurement errors with MeerKAT (grey) and SKA1 (cyan). The lower panel shows the cumulative signal-to-noise (S/N) ratio. }
\label{fig:CHIkIM}
\end{figure}

\section{Cross-correlating with galaxy surveys}
\label{sec:cross}

As we saw above, the prospects for detecting the HI density fluctuations are very good even for a near-term instrument such as MeerKAT-16; however, the measurement of convergence with HI intensity mapping might require an advanced SKA measurement. We also showed that cross-correlating the density and convergence using an IM survey can greatly improve the signal-to-noise ratio for the lensing detection. It is interesting to examine to what extent the HI detections could be accelerated by cross-correlating these measurements with density and convergence derived from galaxy surveys, where the noise and potential systematics are expected to be independent.

For the purposes of these projections we assume that the galaxy power spectrum is related to the density by $P_{gg}(k,z) = b^2_g P_{\rm \delta \delta}(k,z)$ and assume the galaxy bias $b_g(z)$ evolves as $\sqrt{1+z}$ \citep{Rassat:2008ja}.
In addition, there is potential stochasticity between the dark matter density and the galaxy density fields; this is quantified by the correlation coefficient $r_g$.

\subsection{$\delta_{\rm HI} \times \delta_g$ with MeerKAT-16 and DES}

The $\delta_{\rm HI} \times \delta_g$ combination, i.e. the cross-correlation between a 21 cm intensity map with large-scale structure traced by galaxies has been investigated previously \citep{Chang:2010jp,Masui:2012zc} at redshift $z \sim 1$. This correlation constrains $\Omega_{\rm HI}b_{\rm HI}r_{\rm HI-g}$.
For this cross-correlation power spectrum 
\be
P_{\rm HI,g}(k)=\bar{T}b_{\rm HI}b_gr_{\rm HI-g}P_{\delta \delta}(k),
\ee
the uncertainty averaged over a radial bin in $k$-space of width $\Delta k$ is
\begin{align} \nonumber
& \delta P_{\rm HI,g}=\sqrt{2\frac{(2\pi)^3}{V_{\rm sur}}\frac{1}{4\pi k^2 \Delta k}} \times  \\
&\sqrt{P^2_{\rm HI,g}+(P_{\rm HI}+\sigma^2_{\rm pix}V_{\rm pix}W^{-2})(P_{gg}+P^{\rm shot})},
\end{align} 
where the HI noise and shot noise terms were defined above.
{black
For our forecasts here we will set $r_{\rm HI-g}=1$ for simplicity. We will assume MeerKAT-16 measurements with $A_{\rm sky}=5000 \, {\rm deg}^2$ and a total observing time of $1$ week for this case, and combine it with DES. 
The redshift bin we use is $0<z<0.2$ with central redshift $z_c=0.1$. As we can see from Fig.~\ref{fig:PHIg}, these measurements are very precise across a wide range of scales even if a single week's observing time is used.  

We will also be able to perform tomographic studies across the redshift range $0<z<1.45$, constraining the $\Omega_{\rm HI}b_{\rm HI}r_{\rm HI-g}$ combination as a function of redshift. {\color{black} As we discussed in the introduction, constraining $\Omega_{\rm HI}$ is very important for exploiting the power of intensity mapping surveys for cosmology. Performing the aforementioned tomographic studies would measure the late-time evolution of the HI parameters which determine the overall amplitude of the HI signal and, consequently, the signal-to-noise ratio of the clustering measurements.  
In \citet{Switzer:2013ewa}, for example, the measurement of $\Omega_{\rm HI}b_{\rm HI}r_{\rm HI-g}$ was taken as a lower bound of $\Omega_{\rm HI}b_{\rm HI}$ and then combined with the upper bound coming from HI autocorrelation measurements to a determination of $\Omega_{\rm HI}b_{\rm HI}$ at $z\sim 0.8$. As mentioned in the same paper, redshift space distortions can be utilised in order to break the degeneracy between the HI bias and HI density parameters \citep{Wyithe:2008th, Masui:2010mp}. 
}

\begin{figure}
\centerline{
\includegraphics[scale=0.6]{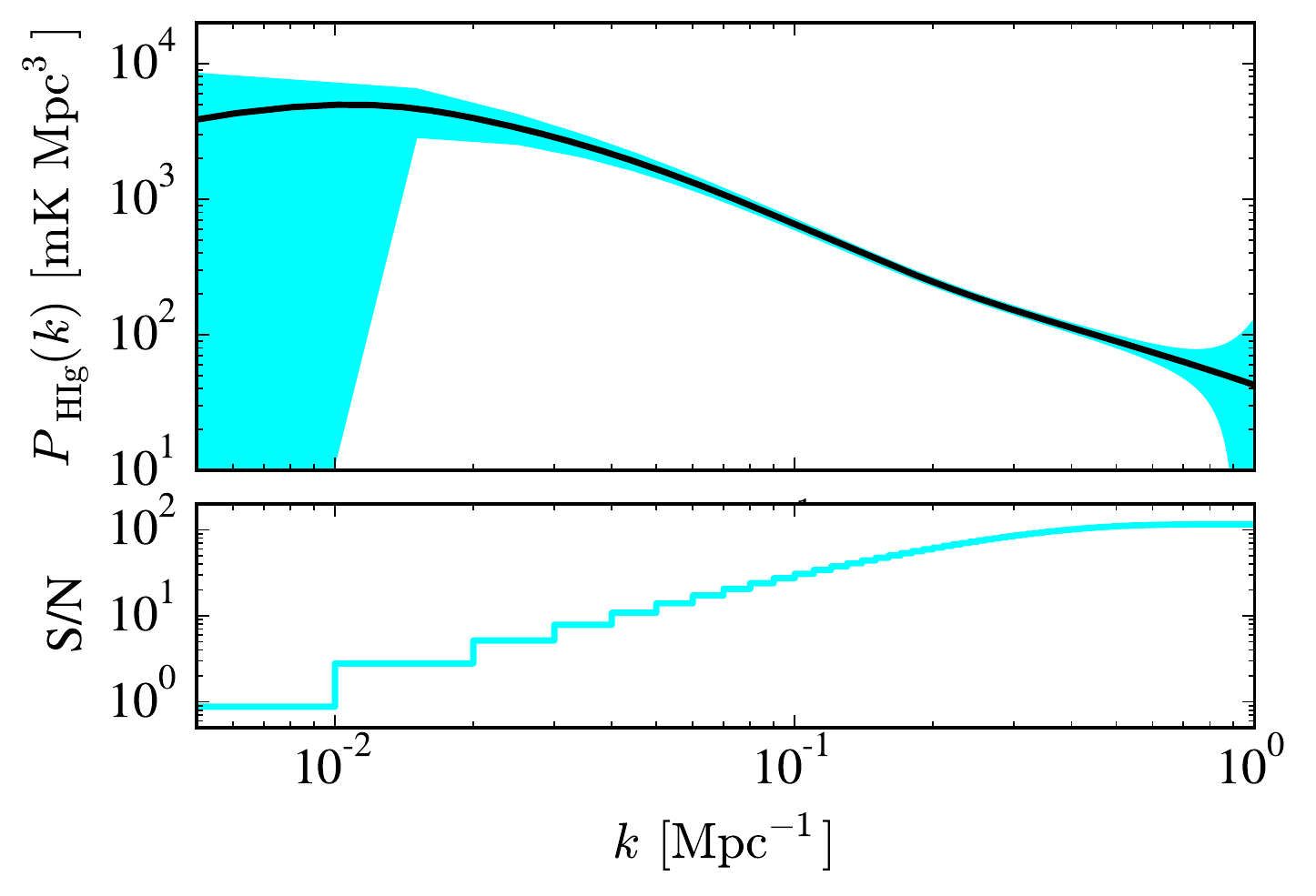}
}
\caption{The upper panel shows the $P_{\rm HI,g}$ cross correlation power spectrum and measurement errors with MeerKAT-16 and DES. The lower panel shows the cumulative signal-to-noise (S/N) ratio. Note $t_{\rm total}=1$ week for MeerKAT-16.}
\label{fig:PHIg}
\end{figure}

\subsection{$\delta_{\rm HI} \times \kappa_g$ with MeerKAT and DES}

We are now going to examine the cross correlation of the HI density fluctuations with the lensing convergence using a galaxy survey. 

The formulae used for the signal and error calculations are the same as in the $\delta_{\rm HI} \times \kappa_{\rm IM}$ case but instead of the IM lensing reconstruction noise $N_\kappa(\ell)$ we have the galaxy survey shape noise $\sigma^2_\kappa/\bar{n}_b$.

For DES lensing measurements, we consider a source bin with $z_b=1.5$ and width $\Delta z = 1.0$. The chosen width contains a large number of galaxies, which translates to a low shape noise in the lensing convergence measurement. As already stated, we always assume $\sigma_\kappa=0.3$ \citep{Schmidt:2011qj}. For MeerKAT we use a bin with central redshift $z_c=0.1$ and width $\Delta z \simeq 0.08$ (equivalently, $\Delta f=100 \, {\rm MHz}$), with $t_{\rm total}=15$ weeks. We also take $A_{\rm sky}=5000 \, {\rm deg}^2$ and $\Delta \ell=50$. The results for MeerKAT-16 are shown in Fig.~\ref{fig:CHIk} ---
note that the dominant noise term is from the HI noise $N(\ell)$ defined in Eq.~(\ref{eq:Nelldish}) which diverges as we reach the limits set by the beam resolution. 

In Fig.~\ref{fig:CHIk_Meer} we show the results for the full MeerKAT, instead in interferometer mode.  One can see that this mode allows smaller scales to be probed with significant signal-to-noise.  
These measurements can constrain the $\Omega_{\rm HI}b_{\rm HI}r_{\rm HI}$ combination. Using tomography this can be achieved across a wide range of redshift, which is very important as there is currently a lot of uncertainty regarding the HI evolution with cosmic time. 

\begin{figure}
\centerline{
\includegraphics[scale=0.6]{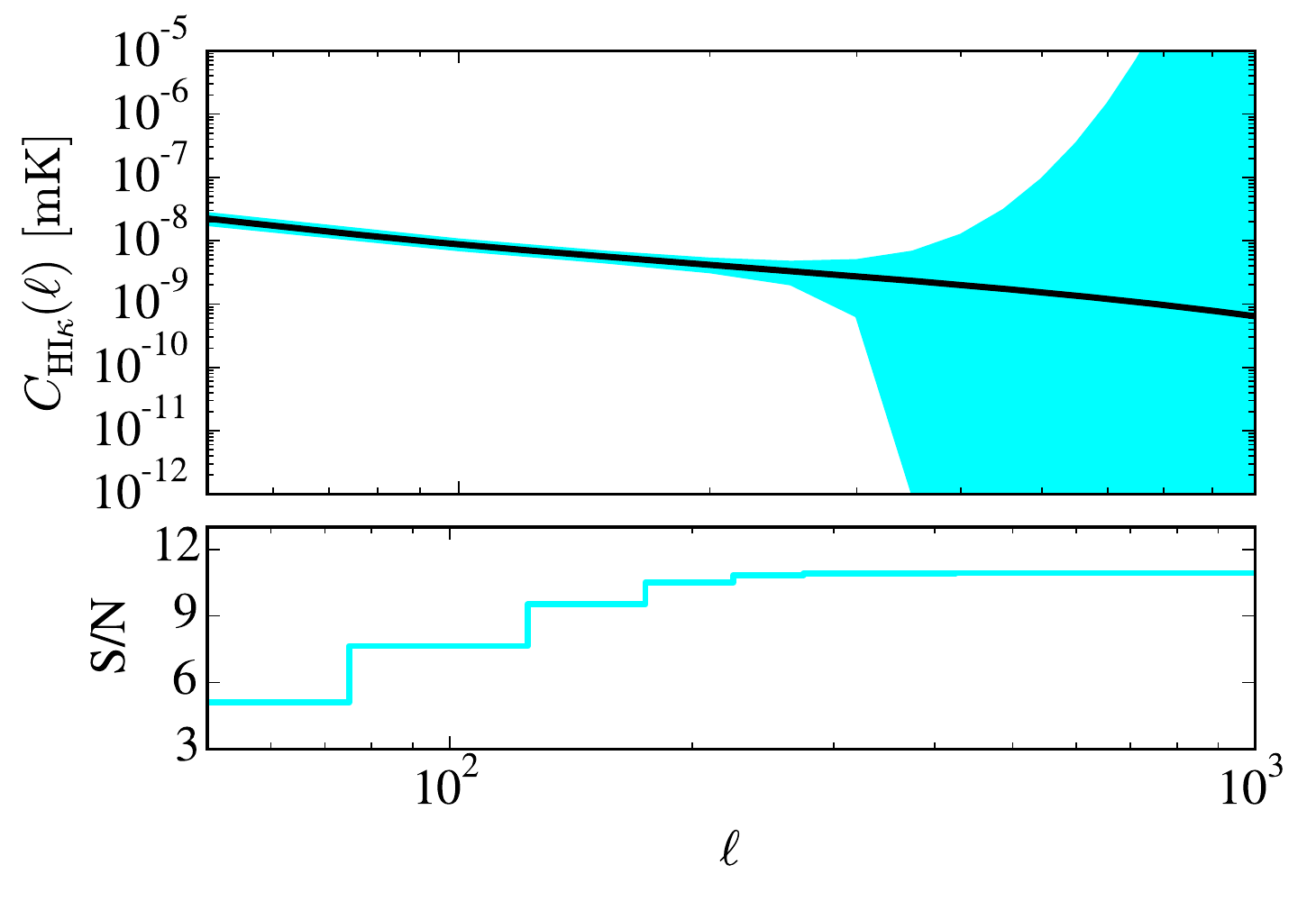}
}
\caption{The upper panel shows the $C_{\rm{HI}\kappa}$ cross correlation power spectrum and measurement errors with MeerKAT-16 and DES. The lower panel shows the cumulative signal-to-noise (S/N) ratio. }
\label{fig:CHIk}
\end{figure}

\begin{figure}
\centerline{
\includegraphics[scale=0.6]{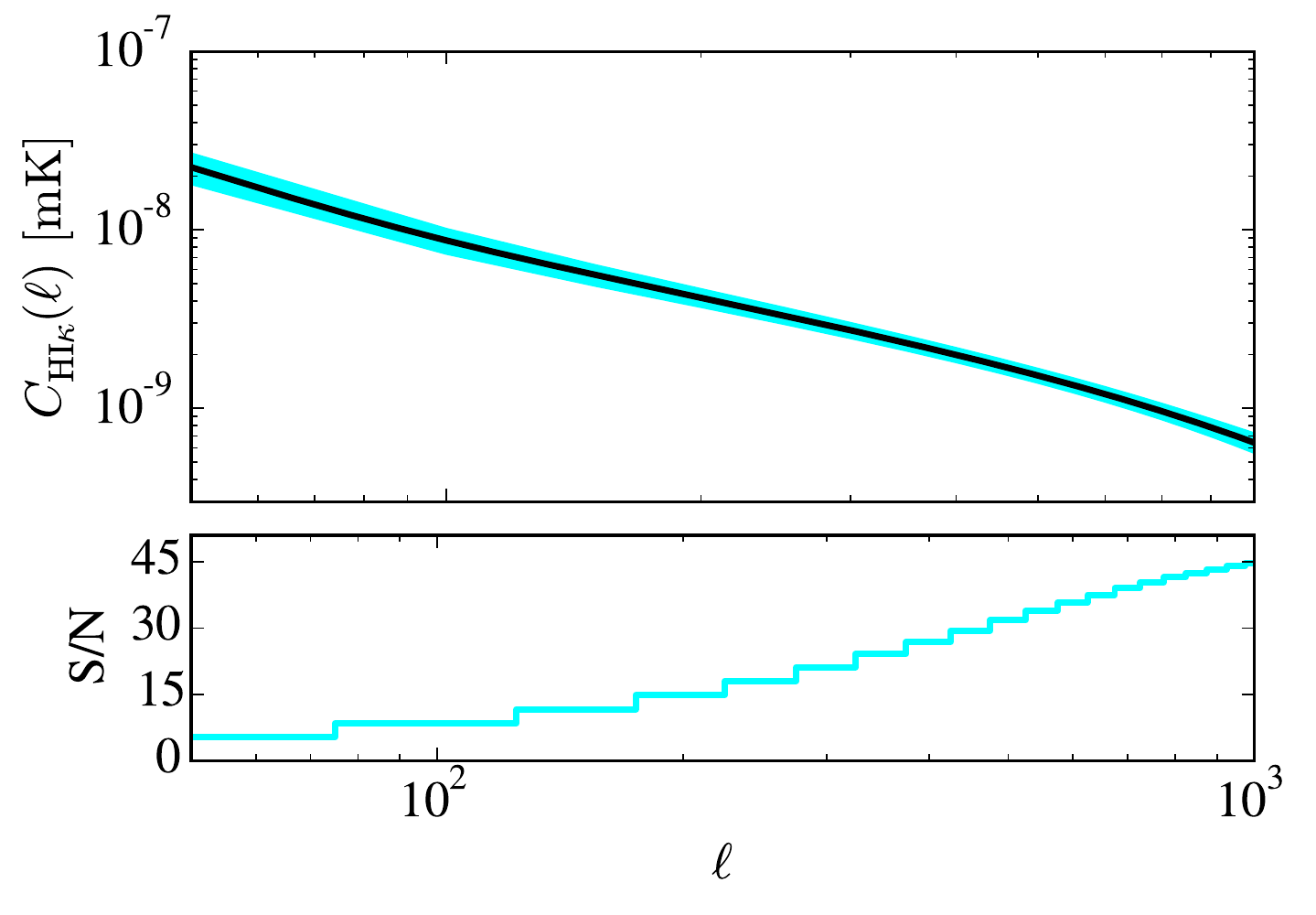}
}
\caption{The upper panel shows the $C_{\rm{HI}\kappa}$ cross correlation power spectrum and measurement errors with MeerKAT (in interferometer mode) and DES. The lower panel shows the cumulative signal-to-noise (S/N) ratio. }
\label{fig:CHIk_Meer}
\end{figure}

\subsection{$\delta_g \times \kappa_{\rm IM}$ with LSST and MeerKAT/SKA1}

A very interesting combination to consider is the cross-correlation of the galaxy density field with the lensing convergence probed via the intensity mapping method. Cross-correlating $\kappa_{\rm IM}$ with $\delta_g$ can help boost the signal-to-noise ratio of the $\kappa$ detection using the method developed in \citet{Pourtsidou:2014pra} and also remove systematic effects since optical and intensity mapping surveys use completely different instruments and strategies.
In this case we have
\be
C_{\rm g \kappa}(\ell)=\frac{3\Omega_{\rm m}H^2_0}{2c^2}\left( \frac{r_g b_g(\hatchi_f) P_{\rm \delta \delta}\left(\frac{\ell}{\hatchi_f},\hatchi_f\right)}{a(\hatchi_f)}\right) \frac{\hatchi_s-\hatchi_f}{\hatchi_f \hatchi_s}.
\ee 
The corresponding uncertainty is
\begin{align} \nonumber
&\delta C_{\rm g\kappa}(\ell) = \sqrt{\frac{2}{(2\ell+1)\Delta \ell f_{\rm sky}}} \times \\
&\sqrt{C^2_{\rm g\kappa}(\ell)+\left(C_{\rm gg}(\ell)+\frac{1}{\bar{n}_g}\right)\left(C_{\kappa \kappa}(\ell)+N_\kappa(\ell)\right)}.
\end{align}
Here 
\be
C_{\rm gg}(\ell) = \int dz E(z) W^2(z) P_{\delta \delta}(\ell/\chi(z),z)/\chi^2(z)
\ee and $\bar{n}_g$ is the number density of galaxies in the redshift bin under consideration.

We show results in Fig.~\ref{fig:Cgk} combining LSST and SKA1, as well as LSST and MeerKAT, with the 21 cm sources at redshift $z_s=1.4$ and the foreground density tracer field at $z_f=1.0$ with $\Delta z_f=0.2$. We use $\Delta \ell =50$.

\begin{figure}
\centerline{
\includegraphics[scale=0.6]{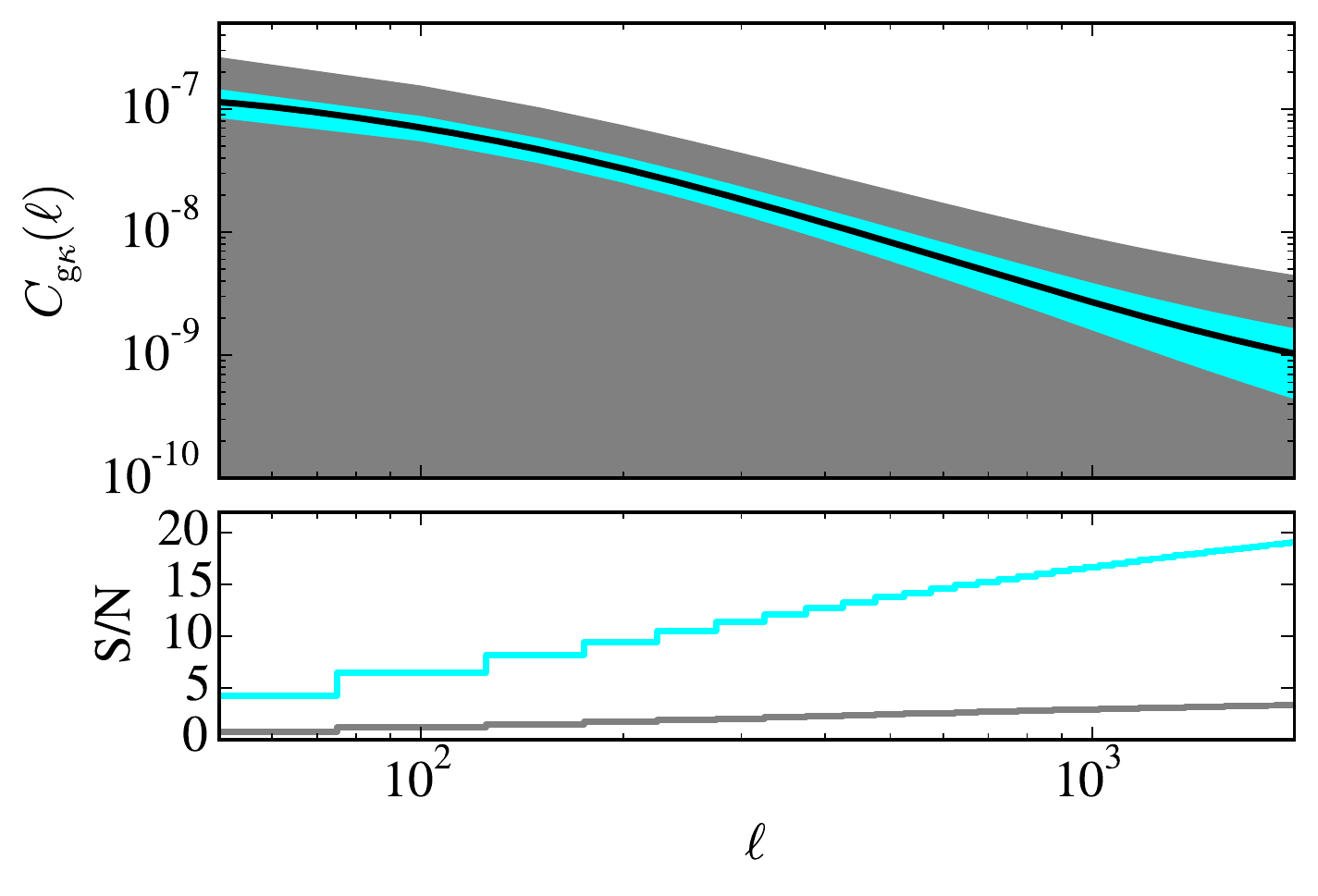}
}
\caption{The upper panel shows the $C_{g\kappa}$ cross correlation power spectrum and measurement errors with LSST and SKA1 (cyan), and LSST and MeerKAT (grey). The lower panel shows the cumulative signal-to-noise (S/N) ratio.}
\label{fig:Cgk}
\end{figure}

{\color{black} We see that a high signal-to-noise detection can be achieved with SKA1 in combination with an optical survey like LSST. Comparing with the HI autocorrelation results presented in Fig.~\ref{fig:CHIkIM}, we see that the $\delta_{\rm HI} \times \kappa_{\rm IM}$ cross-correlation is more powerful; however, the $\delta_g \times \kappa_{\rm IM}$ correlation we considered here is less prone to systematic effects.}

\subsection{$\kappa_g \times \kappa_{\rm IM}$ with LSST and MeerKAT/SKA1}

Finally, we cross-correlate the lensing convergence $\kappa_g$ measured with LSST from sources within our chosen bin centred at $z_b=1$, and $\kappa_{\rm IM}$ measured with the MeerKAT/SKA1 instruments assuming 21 cm sources at $z_s=1.4$.
The results are shown in Fig.~\ref{fig:CkgkIM}. We have used $\Delta \ell = 50$.
\begin{figure}
\centerline{
\includegraphics[scale=0.6]{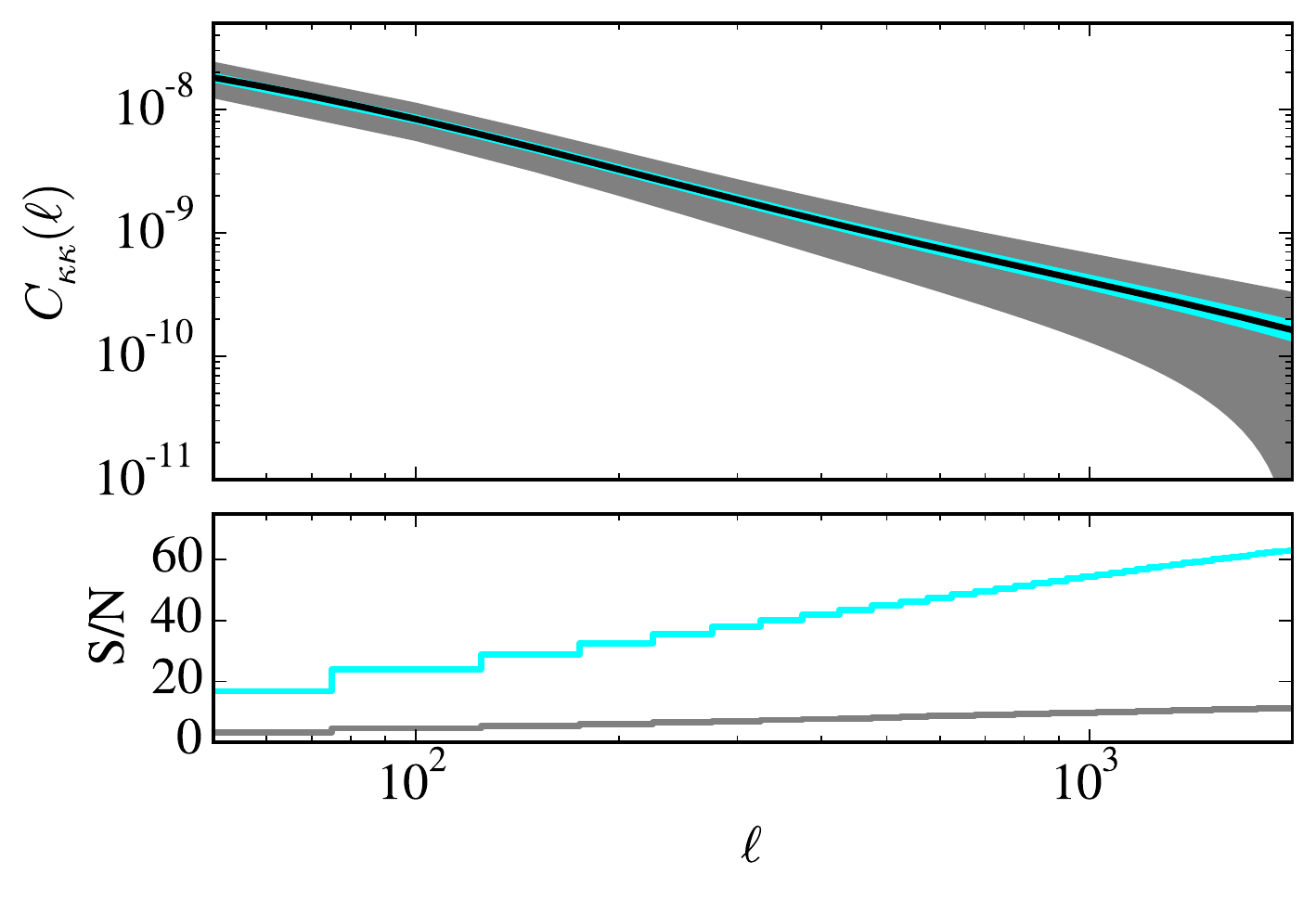}
}
\caption{The upper panel shows the $C_{\kappa \kappa}$ cross correlation power spectrum and measurement errors with LSST and SKA1 (cyan), and LSST and MeerKAT (grey). The lower panel shows the cumulative signal-to-noise (S/N) ratio. }
\label{fig:CkgkIM}
\end{figure}

{\color{black} We see that, in combination with a powerful optical galaxy survey like LSST, both MeerKAT and Phase 1 of the SKA can achieve detection of the lensing convergence coming from 21 cm sources with a high signal-to-noise. This combination could also alleviate issues arising from systematic effects.}

Furthermore, as shown in \citet{Pourtsidou:2014pra}, Phase 2 of the SKA (SKA2) can provide high precision measurements of $\kappa$ (in auto correlation) at redshifts $z \sim 2-3$. This means that we can use tomographic studies along many redshift bins in order to map the evolution of the growth function at redshifts higher than those of galaxy shear surveys. This will be the subject of future work.

\section{Discussion and Conclusions}

In this paper, we have shown how ongoing and future intensity mapping surveys and optical galaxy surveys can be used to perform high precision clustering and lensing measurements. We considered a range of HI surveys, concentrating on the performance of the MeerKAT SKA pathfinder, as well as the full SKA Phase 1, and the DES and LSST optical galaxy surveys. 

Our auto correlation forecasts show that high signal-to-noise HI detection can be achieved already with the first phase of MeerKAT, MeerKAT-16. This is very important for testing the intensity mapping method and calibrating the HI evolution across redshift using tomographic measurements from MeerKAT and Phase 1 of the SKA.  

The measurement of the lensing convergence in auto correlation is much more demanding and heavily depends on the unknown evolution of the HI density \citep{Pourtsidou:2014pra}. Our cross correlation studies show that using the HI or galaxy density fields in cross correlation with $\kappa_{\rm IM}$ considerably improves the 21 cm lensing detection prospects. The same is true when using $\kappa_g$ in cross correlation with $\kappa_{\rm IM}$. Cross-correlating the galaxy and HI densities will also give us information about the galaxy-HI correlation coefficient.  A significant advantage of cross correlating HI intensity mapping and optical galaxy surveys is the alleviation of the issues arising from systematic effects. 

The prospects of detecting -for the first time- HI clustering and lensing of 21 cm emission using the intensity mapping technique with the MeerKAT pathfi\cite{Masui:2009cj,Masui:2010mp,Wyithe:2008th,Sanchez-Ramirez:2015adp,Bigot-Sazy:2015jaa,ECPauto,Amendola:2012ys,Zaldarriaga:2003du,Smith:2002dz,Abell:2009aa,Becker:2015ilr,Crighton:2015pza,Camera:2013kpa,Abate:2012za,Zwaan:2003hp,Santos:2015bsa,Masui:2012zc,Chang:2010jp,Bull:2014rha,Switzer:2013ewa,Battye:2012tg,Ansari:2011bv,Seo:2009fq,Wyithe:2007rq,Loeb:2008hg,Mao:2008ug,Chang:2007xk,McQuinn:2005hk,Battye:2004re,Peterson:2009ka,Dewdney13,Limber:1954zz,Peroux:2001ca,Pourtsidou:2013hea,Pourtsidou:2014pra}nder are particularly exciting. HI can be detected with high signal-to-noise ratio with MeerKAT-16, which is expected to be commissioned in 2016. Using the full MeerKAT instrument in interferometer mode --expected 2017/18-- we have the possibility of detecting 21 cm lensing using IM. This will be an important science achievement of the method and will give us valuable information on how to exploit it for higher redshifts using SKA1. 

{\color{black} Clustering and lensing measurements performed using the intensity mapping technique with SKA1 and its pathfinders, as well as cross-correlations with optical galaxy surveys, have a wide range of further cosmological applications. SKA1-MID can measure redshift space distortions across a wide range of redshift ($0\leq z \leq 2.5$) and is competitive with galaxy surveys like Euclid \citep{Raccanelli:2015hsa}. An intensity mapping survey with SKA1-MID can also constrain primordial non-Gaussianity with $\sigma_{f_{\rm NL}}=2.3$, which is much better than current Planck constraints \citep{Santos:2015bsa}. In \citet{Bull:2015lja}, it was shown that SKA1 IM surveys can yield sub-1\% measurements of the linear growth rate, $f\sigma_8$, for $z \leq 1$. The possibility of testing General Relativity at large scales using HI intensity mapping and optical surveys (in combination with CMB lensing surveys) and the $E_{\rm G}$ statistic was investigated in \citet{Pourtsidou:2015ksn}, showing that sub-1\% $E_{\rm G}$ measurements can be achieved. IM observations can be also used to constrain neutrino masses \citep{Villaescusa-Navarro:2015cca}. } 

Finally, we note that in future work we plan to extend these studies to include forecasted constraints on the HI density $\Omega_{\rm HI}$, the HI bias $b_{\rm HI}$, the galaxy-HI correlation coefficient $r_{\rm HI-g}$ and other cosmological parameters.

\section{Acknowledgments} 
This work was supported by STFC grant ST/H002774/1. RBM's work is  part of the project GLENCO, funded under the Seventh Framework Programme, Ideas, Grant Agreement n. 259349. The authors would like to thank Philip Bull, Stefano Camera, Roy Maartens and Mario Santos for useful discussions and feedback. 

\bibliographystyle{mn2e}
\bibliography{references_21cm}

\end{document}